\\

Title: Determination of the efficiency of a detector in gamma spectrometry of large-volume samples

Authors: E. G. Tertyshnik and A. T. Korsakov

The experimental – calculated method is proposed to determine the full energy peak efficiency (FEPE) of detectors $\varepsilon(E)$ in case a measurement of the large-volume samples. Water is used as standard absorber in which the linear attenuation coefficients for photons $\mu_0(E)$ is well known. The value $\mu(E)$ in sample material (matrix of the sample) is determined experimentally by means of spectrometer. The formulas are given for calculation of the ratio $\varepsilon(E)/\varepsilon_0(E)$, where $\varepsilon_0(E)$ is FEPE of the detector for photons those are arising in the container filled with water (it is found by adding in the container of the Reference radioactive solutions).

To prove the validity of the method ethanol (density 0,8 g/cm$^3$) and water solutions of salts (density 1,2 and 1,5 g/cm$^3$) were used for simulation of the samples with different attenuation coefficients. Standard deviation between experimental and calculated efficiencies has been about 5 %.

# Determination of the efficiency of a detector in gamma spectrometry of large-volume samples

E. G. Tertyshnik and A. T. Korsakov

Measurements on large volume (up to 1000 ml) preparations in the analysis of environmental samples and samples of rocks is an important problem of applied gamma spectrometry, whose accuracy is in many ways determined by the technique used. The use of the method of direct comparison with a standard presumes the existence of reference or standard preparations, containing a known amount of gamma-emitting nuclides, uniformly distributed in a non-active inert filler (matrix). The standard preparation must exactly model the sample both with respect to geometrical dimensions and with respect to the coefficient of attenuation of γ- radiation in the filler material. The technique used to prepare standard preparations is quite complicated, since it is necessary to ensure a uniform distribution of the introduced radioactive substance by repeated mixing while avoiding uncontrollable losses of radionuclides [1]. In addition, the samples to be analyzed have a different chemical composition and are distinguished by their bulk density (samples of ion-exchange resin, soil samples, ashes from plants, etc.). For this reason, the coefficients of attenuation of radiation in the sample and in the sample preparation are practically never equal, and the reliability of the results obtained by the method of direct comparison with a standard is lowered. In many laboratories γ-spectrometric analyses are performed with the help of detectors whose efficiency is known.

The calibration of detectors over a wide range of γ-quanta energies is performed by filling measuring containers of standard sizes with water and adding Certified reference solutions (CRS) to the water. Since the mass of the added CRS is determined on analytical scales and an ideal distribution of radionuclides over the container volume is ensured in the water, the detection efficiency is measured with high accuracy. By detection efficiency of γ quanta we mean the ratio of the number of counts recorded under the total absorption peak during a chosen interval of time to the number of γ quanta arising in the preparation over the same time interval. The number of γ quanta arising in the preparation is calculated starting from the mass of the sample solution introduced into the container and the data on the certificate accompanying the CRS.

The calibration curves obtained can be used to calculate the content of radionuclides in the soil samples, bottom deposits, etc., if the radionuclides being determined emit hard γ- rays (for example, cesium 137), since it is known that for high γ-ray energies the self-absorption of radiation is approximately the same in water and in the sample material. In the general case, the efficiency $\varepsilon(E)$ of detection of γ- radiation from the sample, placed in a standard container, differs from the values $\varepsilon_0(E)$ which are obtained for the container filled with water with known specific activity. This difference is all the more noticeable the larger the difference between the attenuation coefficients of the sample materials and the water.

In this paper we examine a method which permits determining the coefficient $\omega(E) = \varepsilon/\varepsilon_0$ and, therefore, calculating the function $\varepsilon(E)$, if the dependence $\varepsilon_0(E)$ for water has been obtained, for a sample with arbitrary density and unknown chemical composition. Coefficient $\omega$ is a function of the geometrical dimensions of the preparation and of the detector, their mutual arrangement, and also the attenuation coefficients of the sample material and the water. The flux density of $\gamma$-rays for volume sources with different configuration was calculated, taking into account self-absorption, in [2] by integrating the effects produced by point sources taking into account the attenuation kernel (the influence function of a point source). In [3] the results of such calculations were used to estimate the influence of variations in the density of the measured samples on the characteristics of the $\gamma$-spectrometric setup with a small number of channels.

For a cylindrical source with a radius R and height H the flux density of unscattered $\gamma$-rays F (cm$^{-2}$·sec$^{-1}$) at the center of its base is equal to

$$F = S G(\mu H, \mu R) / 2\mu$$

where S is the specific yield of the volume source, i.e., the number of quanta emitted per unit volume per unit time, cm$^{-3}$·sec$^{-1}$; $\mu$ is the linear coefficient of attenuation of radiation, cm$^{-1}$; G is a special function whose values are found from the curves in [2, 4]. Therefore, for a cylindrical container the coefficient $\omega$ can be determined from the following formula

$$\omega = \mu_0 G(\mu H, \mu R) / \mu G(\mu_0 H, \mu_0 R) \qquad (1)$$

A source in the form of a hemisphere with radius L creates at the center of its base a $\gamma$-ray flux density of

$$F = S[1 - \exp(-\mu L)] / 2\mu$$

For this reason, if the measured preparation is shaped like a hemisphere, then the coefficient $\omega$ is calculated using formula

$$\omega = \mu_0 [1 - \exp(-\mu L)] / \mu [1 - \exp(-\mu_0 L)] \qquad (2)$$

The same formula is valid for a preparation in the form of a spherical layer with a thickness L with a point detector placed at the geometric center of the sphere. We note that in the latter case the value of $\omega$ does not depend on the radius of the sphere. A container with a complex shape, with whose help the "well in the sample" geometry is realized (Fig. 1), can be viewed as a combination of two cylindrical sources, and the source of height h and radius r should be viewed as a source with a negative intensity. For such a container the value of ω is found from the formula

$$\omega = \mu_0 [G(\mu H, \mu R) - G(\mu h, \mu r)] / \mu [G(\mu_0 H, \mu_0 R) - G(\mu_0 h, \mu_0 r)] \qquad (3)$$

The value of the coefficient of linear attenuation of radiation, which is necessary in calculating to, can be measured quite simply for any bulk material or liquid. It is known that the flux density of unscattered γ- quanta behind the absorbing layer for a diverging beam of radiation from a point source is determined from the formula

$$F = F_0 b^2 \exp(-\mu t) / (b + t)^2,$$

where $F_0$ is the flux density of γ- quanta on the surface of the absorber, F is the flux density of quanta of unscattered radiation after passage through an absorber of thickness t and **b** is the distance from the source to the surface of the absorber (cm).

Since at the peak of total absorption the detector registers unscattered γ- quanta, and the coefficient of linear attenuation for air is small (for example, for 60-keV quanta $\mu = 0,00022$ cm$^{-1}$ [5] ), when determining the coefficient of linear attenuation of radiation of any substance it is sufficient to:

fix the source, emitting quanta with energy $E_1$, relative to the detector; measure the counting rate under the corresponding total-absorption peak with an empty container (a layer of air) and with the absorber placed into the container; calculate the value of $\mu(E_1)$ from the relation

$$\mu(E_1) = \lg(\mathbf{n/m}) / \mathbf{t} \qquad (4)$$

where $\mu(E_1)$ is the linear coefficient of attenuation (cm$^{-1}$) of γ-rays with energy $E_1$; **t** is the thickness of the layer of absorbing material in the container (cm), **n** is the counting rate with an empty container, and **m** is the counting rate with the absorber or sample placed into the container.

As sources of radiation, it is convenient to use sources from the set of Standard spectrometric γ- sources (SSGS). They are fixed with the help of a centering insert, placed on the top of the measurement container, as shown in Fig. 1. If the coefficient of attenuation is being measured for active samples, then **m** in formula (4) must be replaced by **k** = **m** – **f**, where f is the counting rate under the total-absorption peak corresponding to $E_1$ arising due to the radioactivity of the sample. When measuring **f**, the source 4 (see Fig. 1) must be removed.

The suitability of the proposed method for calculating the efficiency was checked on a γ-spectrometric setup with a DGDK-80B Ge(Li) detector for two containers with volumes of 500 and 1000 cm$^3$. The diameter of the cryostat of the detector was equal to 90 mm. The values of μ were found with the help of the formula (4) for aqueous solutions of cadmium sulfate with a density of 1,2 and 1,5 g/cm$^3$ and ethanol with a density of 0.8 g/cm$^3$. These solutions simulated samples with different attenuation coefficients. We used the numerical values of μ, characterizing solutions with different density, to calculate the coefficients ω according to formulas (l)-(3). It turned out that the values of to obtained using formula (2) for both types of containers differ to a lesser extent from the experimental data than the values calculated using formula (1) or (3). In addition, formulas (1) and (3) are inconvenient for practical applications, because the function G(μH, μR) is not tabulated and determining the values of this function from graphs is a time-consuming process and introduces additional errors. The function Z(μH, μR, R/H, a/H), which describes the radiation field of a cylindrical volume source at a distance *a* from its end face, turned out to be unsuitable for calculating ω [4].

The values of ω for a complex-shaped' container with a volume of 1000 cm$^3$, calculated using formula (2), are presented in Table 1. The table also gives the values of $\varepsilon/\varepsilon_0 = \Omega$ obtained experimentally. We performed the experiment as follows. We added several milliliters of the solution of radionuclide (for example, $^{241}$Am) into the container with the absorber modeling the sample. After carefully mixing the contents of the container, we measured the counting rate under the corresponding total-absorption peak (60 keV) and scaled it to the volume of- the solution Introduced, finding the value of N, pulses/(s·ml). We then determined the counting rate $N_0$, normalized to the volume of the solution introduced, with a container filled with water, tagged with the same radioactive solution, placed on the detector. Since the normalized counting rate is proportional to the detection efficiency,

$$\Omega = N/N_0.$$

In experiments of this type, powdered materials are usually used to model the sample: soda, quartz sand, chromium oxide, etc. [3]. The use of salt solutions in containers as absorbing media increases the accuracy with which $\varepsilon/\varepsilon_0$ is determined because of the uniform distribution of the radionuclides introduced over the volume of the container.

It follows from the table that for ethanol in the energy range 60-835 keV the computed and experimental values of w are close to one. For solutions of cadmium sulfate the rms relative deviation of the computed values from the data obtained by model experiments is equal to 5%. The maximum deviation does not exceed 10%. Analogous results were obtained for a cylindrical container with a volume of 500 cm$^3$ (Fig. 2). Taking into account the spread of the experimental data and the error in the determination of the coefficients of linear attenuation, it is evident that the computer and experimental values are in good agreement.

It has thus been shown that in order to determine the efficiency of detection of γ- radiation by a detector when measuring samples with a large volume, the following are necessary:

using standard radioactive solutions, the dependence $\varepsilon_0(E)$ must be obtained with the measuring container filled with water;

the coefficient of linear attenuation of the sample material (matrix) must be determined using formula (4); the value of $\mu_0(E)$ for water can be taken from the tables [5, 6];

the efficiency of the detector for radiation from the sample must be calculated using formula $\varepsilon(E) = \omega\, \varepsilon_0(E)$, finding the value of the coefficient $\omega$ for the relation (2).

For a cylindrical container $l = H$ and for a complex-shaped container $l = H - h$.

The activity of the radionuclide, emitting γ- rays with energy $E_1$ is calculated from the relation

$$A = F/T\acute{\eta}\,\omega\,\varepsilon_0(E_1),$$

where A is the activity of the radionuclide in the sample in Bq; F is the area of the total absorption peak in counts; T is the duration of the measurements in sec; and $\acute{\eta}$ is the quantum yield of the radionuclide.

Compared with the method of calibration of the spectrometer based on the efficiency, which is proposed in [7] and recommended for wide application [8], the method described here is distinguished by its universality and lower labor intensiveness.

## LITERATURE CITED

1. Ts. I. Bobovnikova, S. B. Iokhel'son, and V. N. Churkin. In: Apparatus and Methods for Studying Environmental Pollutants [in Russian], No. 2, Gidrometeoizdat, Leningrad (1970), p. 117.
2. Shielding of Nuclear Reactors [Russian^translation]. IL, Moscow (1958).
3. V. I. Parkhomenko, Ë*. M. Krisyuk, and E. P. Lisachenko. Prib. Tekh. Eksp., No. 4, 46 (1983).
4. Handbook on Radiation Shielding for Engineers [in Russian]. Atomizdat, Moscow (1973).
5. E. Storm, H. Israel. Photon cross sections from 0,001 to 100 MeV for elements 1 through 100.

Los Alamos Laboratory, New Mexico, 1967, 256 p.

6. N. H. Cutshall, I. L. Larsen, C. R. Olsen. Direct analysis of $^{210}$Pb in sediments: Self- absorption corrections. / Nuclear Instruments and Methods in Physics Research, Vol. 206, issues 1 – 2, 1983, p.309-312.

7. G.G. Doroshenko et al. in Problems of Ensuring Radiation Safety during Operation of Nuclear Power Plants [ in Russian], Vol. 2, Prague (1976), p. 24.

8. Methodological Recommendations on Public-Health Monitoring of the Content of Radioactiv Materials in Objects in Environment [in Russian], Minzdrav USSR, Moscow (1980).

Table 1. Coefficients $\varepsilon/\varepsilon_0$ Calculated ($\omega$) Using Formula (2) and Obtained Experimentally ($\Omega$) for a Container with a Volume of 1000 cm$^3$ *(l = H - h)*

| Energy of γ –rays, keV | Density, g/cm$^3$ | | | | | |
|---|---|---|---|---|---|---|
| | 1,5 | | 1,2 | | 0,8 | |
| | $\omega$ | $\Omega$ | $\omega$ | $\Omega$ | $\omega$ | $\Omega$ |
| 60 | 0,18 | 0,18 | 0,38 | 0,37 | 1,1 | 1,0 |
| 88 | 0,44 | 0,45 | 0,67 | 0,71 | 1,05 | 1,1 |
| 122 | 0,64 | 0,65 | 0,79 | 0,80 | 1,03 | 1,03 |
| 135 | 0,71 | 0,72 | 0,86 | 0,85 | 1,05 | 1,01 |
| 166 | 0,78 | 0,82 | 0,91 | 0,91 | 1,02 | 1,02 |
| 320 | 0,93 | 0,98 | 0,94 | 1,02 | 1,05 | 1,09 |
| 392 | 0,92 | 0,97 | 0,96 | 1,06 | 1,02 | 1,05 |
| 662 | 0,95 | 0,92 | 0,98 | 0,98 | 1,03 | 0,97 |
| 835 | 0,95 | 0,91 | 0,99 | 0,97 | 1,02 | – |

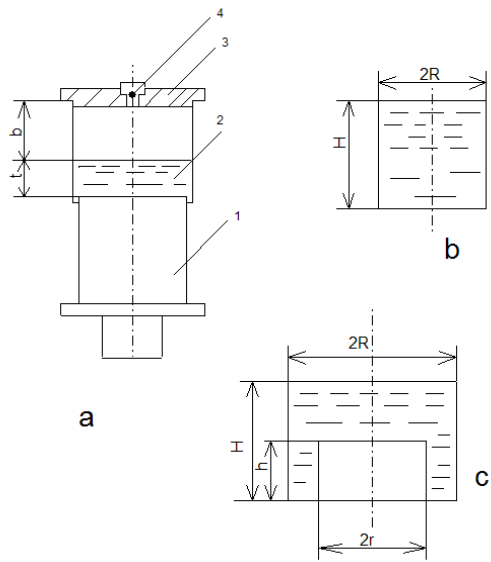

Fig. 1. Measurement geometry:

a) scheme for determining the coefficient of linear attenuation of radiation by the sample material:

1) detector; 2) sample; 3) stabilizing insert; 4) radiation source;

b) cylindrical container with a volume of 500 cm$^3$ (H = 53 mm, R = 55 mm);

c) complex-shaped container with a volume of 1000 cm$^3$ (H = 87 mm, R = 71 mm, h = 52 mm, r = 47 mm, the inner dimensions are indicated).

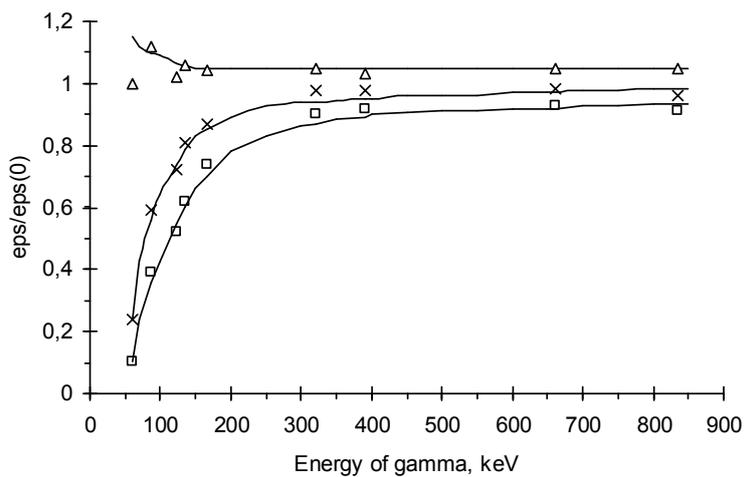

Fig. 2. Energy dependence of the ratio $\varepsilon/\varepsilon_0$ for materials with different density:

∆) ethanol; ×) solution density of 1,2 g/cm$^3$; □) solution density of 1,5 g/cm$^3$.